\documentclass[prd,preprint,nofootinbib,superscriptaddress]{revtex4}
\usepackage[latin9]{inputenc}
\usepackage{amsmath}
\usepackage{amssymb}
\usepackage{graphicx}
\usepackage{color}

\makeatletter
\@ifundefined{textcolor}{}
{%
 \definecolor{BLACK}{gray}{0}
 \definecolor{WHITE}{gray}{1}
 \definecolor{RED}{rgb}{1,0,0}
 \definecolor{GREEN}{rgb}{0,1,0}
 \definecolor{BLUE}{rgb}{0,0,1}
 \definecolor{CYAN}{cmyk}{1,0,0,0}
 \definecolor{MAGENTA}{cmyk}{0,1,0,0}
 \definecolor{YELLOW}{cmyk}{0,0,1,0}
}

\makeatother

\begin{document}
\title{Hubble tension in lepton asymmetric cosmology with an extra radiation }

\author{Osamu Seto}
\email{seto@particle.sci.hokudai.ac.jp}
\affiliation{Institute for the Advancement of Higher Education, Hokkaido University,
Sapporo 060-0817, Japan}
\affiliation{Department of Physics, Hokkaido University, Sapporo 060-0810, Japan}

\author{Yo Toda}
\email{y-toda@particle.sci.hokudai.ac.jp}
\affiliation{Department of Physics, Hokkaido University, Sapporo 060-0810, Japan}

\begin{abstract}
%%%%%%%%%%%%%%%%%%%%%%
We study the fit of cosmological models with two additional free parameters $N_\mathrm{eff}$ and $\xi_e$ in addition to the parameters of $\Lambda$CDM. We introduce extra radiation components such as hot axions or sterile neutrinos in addition to the energy density of neutrinos with large neutrino degeneracy. Then, a larger $N_\mathrm{eff}$ is allowed without spoiling Big Bang Nucleosynthesis (BBN), 
as positive neutrino degeneracy $\xi_e$ could improve BBN fit. By analysing the data from Planck, baryon acoustic oscillation (BAO), BBN and type-Ia supernovae (SNeIa), it can be seen that the Hubble tension can be ameliorated for $\xi_{e}\simeq 0.04$ and $0.3 \lesssim \Delta N_\mathrm{eff} \lesssim 0.6$.
%%%%%%%%%%%%%%%%%%%%%%
\end{abstract}
\preprint{EPHOU-21-008}

\maketitle
%\vspace*{1cm}

%\makeatletter

%%%%%%%%%%%%%%%%%%%%%%%%%%%%%% LyX specific LaTeX commands.
%% Because html converters don't know tabularnewline
%\providecommand{\tabularnewline}{\\}

%\makeatother

%\begin{document}

\section{Introduction}

%%%%%%%%%%%%%%%%%%%%%%%

The effective number of relativistic degrees of freedom, $N_\mathrm{eff}$, is 
one of the simplest extension of the standard cosmology and 
effectively constrains an extension of the standard model (SM) of particle physics.
Since the standard model of cosmology, the $\Lambda$CDM model with the cosmological 
constant $\Lambda$ as the dark energy and cold dark matter (CDM), has been successful,
$N_\mathrm{eff}$ has been well constrained by many cosmological observations.
Planck (2018)~\cite{Aghanim:2018eyx} has reported the constraint 
as $N_\mathrm{eff}=2.99\pm 0.17 \, (68 \%)$ based on the combined data of 
cosmic microwave background (CMB) and baryon acoustic oscillation (BAO).
The combined data on abundance of light elements synthesized at the Big Bang 
Nucleosynthesis (BBN) and CMB also provide $N_\mathrm{eff}=2.88\pm 0.16 \, (1\sigma)$~\cite{Cyburt:2015mya}.
Those are consistent with the prediction of the SM of particle physics,
 $N_\mathrm{eff}^\mathrm{SM} \simeq 3.046$~\cite{Mangano:2005cc}. 
For recent progress of this calculation, see e.g., 
Refs.~\cite{Escudero:2018mvt,Bennett:2019ewm,Escudero:2020dfa,Akita:2020szl,Bennett:2020zkv}.

Interestingly, a somewhat larger $N_\mathrm{eff}$ is also indicated 
by the discrepancy between the measured values of the Hubble parameter by local 
(low red shift) measurements and that by distant (high red shift) measurements.
Assuming $\Lambda$CDM model, Planck measurements of CMB anisotropy 
infers $H_{0}=67.4\pm 0.5$ km/s/Mpc~\cite{Aghanim:2018eyx}. 
Other distant observations such as the Atacama Cosmology
Telescope~\cite{Aiola:2020azj}, BAO~\cite{Addison:2017fdm}, and the combination
of BAO$+$BBN analysis (independent of CMB)~\cite{Schoneberg:2019wmt} all 
infer $H_{0}\sim 67$ km/s/Mpc. 
On the other hand, local measurements of $H_0$ by the SH0ES collaboration 
with Cepheids and type Ia supernovae (SNe Ia) in Ref.~\cite{Riess:2018uxu} (hereafter, R18) 
and Ref.~\cite{Riess:2019cxk} (hereafter R19)
and by the H0LiCOW collaboration with lensed quasars~\cite{Wong:2019kwg}
have reported as $H_{0} \sim 73$ km/s/Mpc.
Another local measurement using the Tip of the Red Giant Branch (TRGB) 
as distance ladders has obtained a value between Planck and the SH0ES, 
$H_{0} \sim 70$ km/s/Mpc~\cite{Freedman:2019jwv}.

Until now, numerous attempts have been made to this problem. 
See, e.g., Ref.~\cite{DiValentino:2021izs} for a recent review.
One of the simplest approaches is to introduce additional relativistic
degrees of freedom $\Delta N_\mathrm{eff} = N_\mathrm{eff}-N_\mathrm{eff}^\mathrm{SM}$.
This is because that the shorten sound horizon $r_{s*}$ at the recombination epoch 
by the extra energy density with a larger $N_{\mathrm{eff}}$ 
and the measured angular size $\theta_{*}\equiv r_{*}/D_{M*}$ of the acoustic scale 
infers a shorter angular diameter distance $D_{M*} \propto 1/H_0$.
The preferred value has been suggested as $0.2 \lesssim \Delta N_\mathrm{eff}
 \lesssim 0.5$ (CMB$+$BAO$+$R18)~\cite{Aghanim:2018eyx} and 
\begin{equation}
0.2 \lesssim \Delta N_{\mathrm{eff}} \lesssim 0.4, 
\label{Eq:DeltaN-w/o-xi} 
\end{equation}
 (CMB$+$BAO$+$Panthenon~\cite{Scolnic:2017caz}$+$R19$+$BBN)~\cite{Seto:2021xua}, depending upon data sets.
Such an extra $\Delta N_\mathrm{eff}$ can be easily realized in various particle physics 
beyond the SM such as hot axion~\cite{DEramo:2018vss} or the $L_{\mu}-L_{\tau}$ gauge interaction~\cite{Escudero:2019gzq}.
However, as mentioned above, the magnitude of $N_{\mathrm{eff}}$ is constrained 
by various reasons, principally, CMB, BBN and large scale structure of the Universe.
This BBN limit primarily comes from the Helium mass fraction $Y_{P}$ that constrains
 the extra energy density not to speed up the cosmic expansion too much.
Actually, there is a possibility to relax the constraint by $Y_{P}$ significantly.
If our Universe has a large electron-type lepton asymmetry $\xi_{e}$,
 which could suppress the conversion of proton to neutron,
 the resultant $Y_P$ can be consistent with the observation
 for a somewhat large $N_{\mathrm{eff}}$~\cite{BeaudetGoret,Lesgourgues:1999wu,Serpico:2005bc,Shiraishi:2009fu,Kirilova:2013aja,Yang:2018oqg,Caramete:2013bua,Barenboim:2016lxv,Gelmini:2020ekg}. 
Various mechanisms for generating a large lepton asymmetry have been proposed in 
literature~\cite{Foot:1995qk,Shi:1996ic,Casas:1997gx,McDonald:1999in,Kawasaki:2002hq,Yamaguchi:2002vw,Takahashi:2003db,Shaposhnikov:2008pf}.
%Namely, introducing the asymmetry $\xi_{e}$ reinstate $\Delta $
%model as the possible solution.

In this paper, we investigate the fit of cosmological models with additional 
two free parameters $N_\mathrm{eff}$ and $\xi_e$ besides $\Lambda$CDM parameters.
We consider that lepton asymmetry as well as extra radiation components 
in addition to the contribution from neutrino degeneracy given by the following Eq.~(\ref{Neff-xi}) 
are independent free parameters.
This is the essential ingredient in this work.
In fact, previously, it has been pointed out that the $\Delta N_\mathrm{eff}$ from lepton asymmetry 
is not effective to alleviate the Hubble tension in Ref.~\cite{Barenboim:2016lxv}, which is quoted 
at Sec.7.2 in Ref.~\cite{DiValentino:2021izs} as well.
In the paper~\cite{Barenboim:2016lxv}, it has been assumed that all $\Delta N_\mathrm{eff}$ 
are originated from lepton asymmetry and the electron-type lepton asymmetry 
is negligibly tiny to keep BBN intact.
In contrast to those previous works, 
by taking account of the effect by the electron-type lepton asymmetry to the Helium abundance,
we evaluate the fitting to various data, principally including CMB, BAO and BBN and show 
that some amount of electron-type lepton asymmetry of the order of $10^{-2}$ with an extra radiation 
is promising.
%
%In this paper, the combination of local and distant measurements determines
%the best-fit electron lepton asymmetry $\xi_{e}$ to solve the Hubble
%tension. Then, we also estimate Hubble constant at the best-fit $\xi_{e}$
%without local measurements and report that $H_{0}=70$ can be achieved
%in $2\sigma$. 

%This paper is organized as follows. After we review how both $\Delta N_{\mathrm{eff}}$ and
%the degeneracy parameter $\xi_{e}$ affect cosmology in Sec.~\ref{sec:physics}, 
%explain our analysis of data sets in Sec. \ref{sec:analysis}, 
%show our results in Sec. \ref{sec:result}, and
%finally summarize in Sec. \ref{sec:summary}.

%%%%%%%%%%%%%%%%%%%%%%%

\section{Cosmological effects of $N_\mathrm{eff}$ and $\xi_{e}$}
\label{sec:physics}

The extra radiation affects not only CMB but also BBN, 
because the additional component increases the expansion rate of the Universe, 
the decoupling temperature of the weak interaction,
and the neutron-to-proton ratio. 
The increased neutron-to-proton ratio results in the larger helium mass fraction.
The measured Helium mass fraction $Y_{P}=0.2449\pm 0.0040$~\cite{Aver:2015iza}
does not appear to be compatible with a large extra $N_{\mathrm{eff}}$.

To let a larger $N_{\mathrm{eff}}$ available, we introduce non-negligible lepton asymmetry. 
We use the degeneracy parameter 
\begin{align}
\xi_{i} = \frac{\mu_{\nu_i}}{T_\nu} ,
\end{align}
to parameterize the lepton asymmetry, where $\mu_{\nu_i}$ is the chemical potential
 for $i\,(=e,\mu,\tau)$-th flavor neutrino and $T_{\nu}$ is the temperature of neutrinos.
The Fermi-Dirac distribution functions for $i$-th flavor (anti-)neutrinos with non-vanishing
neutrino degeneracy is expressed as
\begin{align}
f_{\nu_{i}}(p,\xi_{i})=\frac{1}{\exp(\frac{p}{T_{\nu}}-\xi_{i})+1}, \qquad f_{\bar{\nu}_{i}}(p,\xi_{i})=\frac{1}{\exp(\frac{p}{T_{\nu}}+\xi_{i})+1}, 
\end{align}
with $p$ being the proper momentum.
The lepton asymmetry $\xi_{i}$ affects both BBN and CMB as (principally)
following three ways~\cite{Lesgourgues:1999wu,Shiraishi:2009fu,BeaudetGoret}.

First, the lepton asymmetry is the difference between the number
of particles and antiparticles. 
From our definition of the lepton asymmetry, 
there are more neutrinos than anti-neutrinos in the positive asymmetric Universe. 
In particular, the electron-type lepton asymmetry
\begin{equation}
L_{e} \equiv \frac{n_{\nu_{e}}-n_{\bar{\nu}_{e}}}{n_{\gamma}}
 = \frac{1}{36\zeta(3)}\left(\frac{T_{\nu_{e}}}{T_{\gamma}}\right)^{3}(\pi^{2}\xi_{\nu_{e}}+\xi_{\nu_{e}}^{3}) ,
\end{equation}
 is interesting. At the beginning of the BBN, neutrons and protons
are in equilibrium % $p+e^{-}\leftrightarrow n+\nu_{e}$
 until the decoupling of the weak interaction. 
If there is sizable positive $\xi_e$, the process $p+\bar{\nu}_{e} \rightarrow n+e^{+}$
 is suppressed compared with the process $n+\nu_{e} \rightarrow p+e^{-}$ because of less anti-electron neutrinos.
%Then, after
%the decoupling of the weak interaction, $\beta$ decay $n\rightarrow p+e^{-}+\bar{\nu}_{e}$
%occurs. Hence, the increase of $\xi_{e}$ and $L_{e}$
With such a suppressed neutron-to-proton ratio, the resulting Helium mass fraction $Y_{P}$ is decreased.
This may compensate for the $Y_P$ increase caused by a larger $N_{\mathrm{eff}}$. 
%to solve the Hubble
%tension and $Y_{P}$ became larger than the observation, we can dispel
%the surplus $Y_{P}$ by introducing the electron asymmetry.
Although we do not deal degeneracy parameters of other flavors $\xi_\mu$ and $\xi_\tau$ 
as explicit input parameters, it is expected that all of those are of 
the same order of magnitudes as $\xi_\mu \sim \xi_\tau \sim \xi_e$ because of neutrino 
oscillation~\cite{Dolgov:2002ab,Wong:2002fa,Abazajian:2002qx,Pastor:2008ti,Barenboim:2016shh}.

Second, the energy density of neutrinos is increased by the neutrino degeneracy as
\begin{align}
\rho_{\nu+\bar{\nu}} &= \sum_{i}T_{\nu}^{4}\int\frac{d^{3}p}{(2\pi)^{3}}p
 \left[ f_{\nu_{i}}(p,\xi_{i})+f_{\bar{\nu}_{i}}(p,\xi_{i})\right] =
\left.\rho_{\nu+\bar{\nu}}\right|_{\xi=0}
 + \Delta\rho_{\nu+\bar{\nu}}(\xi) ,\\
\Delta \rho_{\nu+\bar{\nu}} &\propto T_{\nu}^4 \left( \sum_{\xi_i} \xi_i^{2} +\mathcal{O}(\xi_i^4)  \right) , 
%+\frac{15}{7}\left[\left(\frac{\xi_{e}}{\pi}\right)^{4}+\left(\frac{\xi_{\mu}}{\pi}\right)^{4}+\left(\frac{\xi_{\tau}}{\pi}\right)^{4}\right],
\label{Neff-xi}
\end{align}
with $\left.\rho_{\nu+\bar{\nu}}\right|_{\xi=0}$ being the energy density with vanishing $\xi_i$.
However, for small degeneracy as $\xi<0.1$ as we will consider, the extra contribution 
to the energy density $\Delta\rho_{\nu+\bar{\nu}}(\xi)$ from the neutrino degeneracy $\xi$
is very small for both BBN and CMB. 
The $\xi$ dependent part of $\Delta N_{\mathrm{eff}}$, $\Delta N_{\mathrm{eff}}(\xi)$, 
is $\mathcal{O}(10^{-2})$.  
In any case, given that we are introducing an extra radiation (ER) component, 
one may regard our $N_{\mathrm{eff}}$ as $\Delta N_{\mathrm{eff}} = \Delta N_{\mathrm{eff}}^{\mathrm{ER}}+\Delta N_{\mathrm{eff}}(\xi)$.

%This contribution principally increases $Y_{p}$ and decreases $r_{*}$
%as we discussed above. The energy density of the massive neutrino
%is also affected by leptonic asymmetry. However, the coefficient of
%this modification is $1+\mathcal{O}(10^{-2})$ where $\xi<0.1$ (and
%we consider only this case)~\cite{Shiraishi:2009fu}. Hence, we ignore
%this.

Third, in perturbation level, the asymmetry factor comes into the source term of the Boltzmann equation.
This is given in the synchronous gauge by~\cite{Ma:1995ey},
\begin{equation}
\frac{\partial\Psi}{\partial\tau}+i(\boldsymbol{k}\cdot\hat{n})\Psi 
+\frac{d\ln(f_{\nu}+f_{\bar{\nu}})}{d\ln q}\left[\dot{\eta}-\frac{\dot{h}+6\dot{\eta}}{2}(\hat{k}\cdot\hat{n})^{2}\right]=0 ,
\label{eq:Boltzmann}
\end{equation}
 with $q=ap$.
Here, $\Psi$ is the perturbation to the distribution function, $\hat{n}$ is the direction of the momentum, 
$\tau$ is the conformal time, $\boldsymbol{k}$ is the wave-number
of the Fourier mode, and $h$ and $\eta$ are the synchronous metric perturbations.
In the computation of the power spectrum $C_{l}$, when we integrate
Eq.~(\ref{eq:Boltzmann}) with $(f_{\nu}+f_{\bar{\nu}})q^{3}dq$,  
this $\xi$ dependence in perturbation disappears~\cite{Lesgourgues:1999wu}. 
Accordingly, Eq.~(\ref{eq:Boltzmann}) applies in non-vanishing $\xi$ cases without any change.

\section{Data and Analysis}

\label{sec:analysis}

We perform a Markov-Chain Monte Carlo (MCMC) analysis on a $N_{\mathrm{eff}}$
model with a large lepton asymmetry. 
We use the public MCMC code \texttt{CosmoMC-planck2018}~\cite{Lewis:2002ah}.
For estimation of light elements with non-vanishing lepton degeneracy $\xi_e$, 
we have also used \texttt{PArthENoPE2.0-Standard}~\cite{Consiglio:2017pot}.
Since a small Hubble parameter is indicated if we allow larger neutrino 
masses~\cite{Aghanim:2018eyx}, in order to minimize this effect, 
we assume the smallest neutrino mass with 
the spectrum of normal hierarchy, $m_\nu=(0, 0, 0.06)$ eV, in this paper. 

\subsection{Data sets}

We analyze the models by referring to the following cosmological observation
data sets. We include both temperature and polarization likelihoods
for high $l$ \texttt{plik} ($l=30$ to $2508$ in TT and $l=30$ to $1997$ in
EE and TE) and low$l$ \texttt{Commander} and lowE \texttt{SimAll}
($l=2$ to $29$) of Planck (2018) measurement of the CMB temperature
anisotropy~\cite{Aghanim:2018eyx}. We also include Planck lensing~\cite{Aghanim:2018oex}.
For constraints on low red shift cosmology, we include data of BAO
from 6dF~\cite{Beutler:2011hx}, DR7~\cite{Ross:2014qpa}, and DR12~\cite{Alam:2016hwk}.
We also include Pantheon~\cite{Scolnic:2017caz} of the local measurement
of light curves and luminosity distance of supernovae as well as SH0ES
(R19)~\cite{Riess:2019cxk} of the local measurement of the Hubble
constant from the Hubble Space Telescope observation of Supernovae
and Cepheid variables. Finally, we include the data sets of helium
mass fraction $Y_{P}$ measurement~\cite{Aver:2015iza} and deuterium
abundance $D/H$ measurement~\cite{Cooke:2017cwo} to impose BBN constraints.

\section{Result and discussion}

\label{sec:result}

\begin{figure}[htbp]
\includegraphics[width=14.5cm]{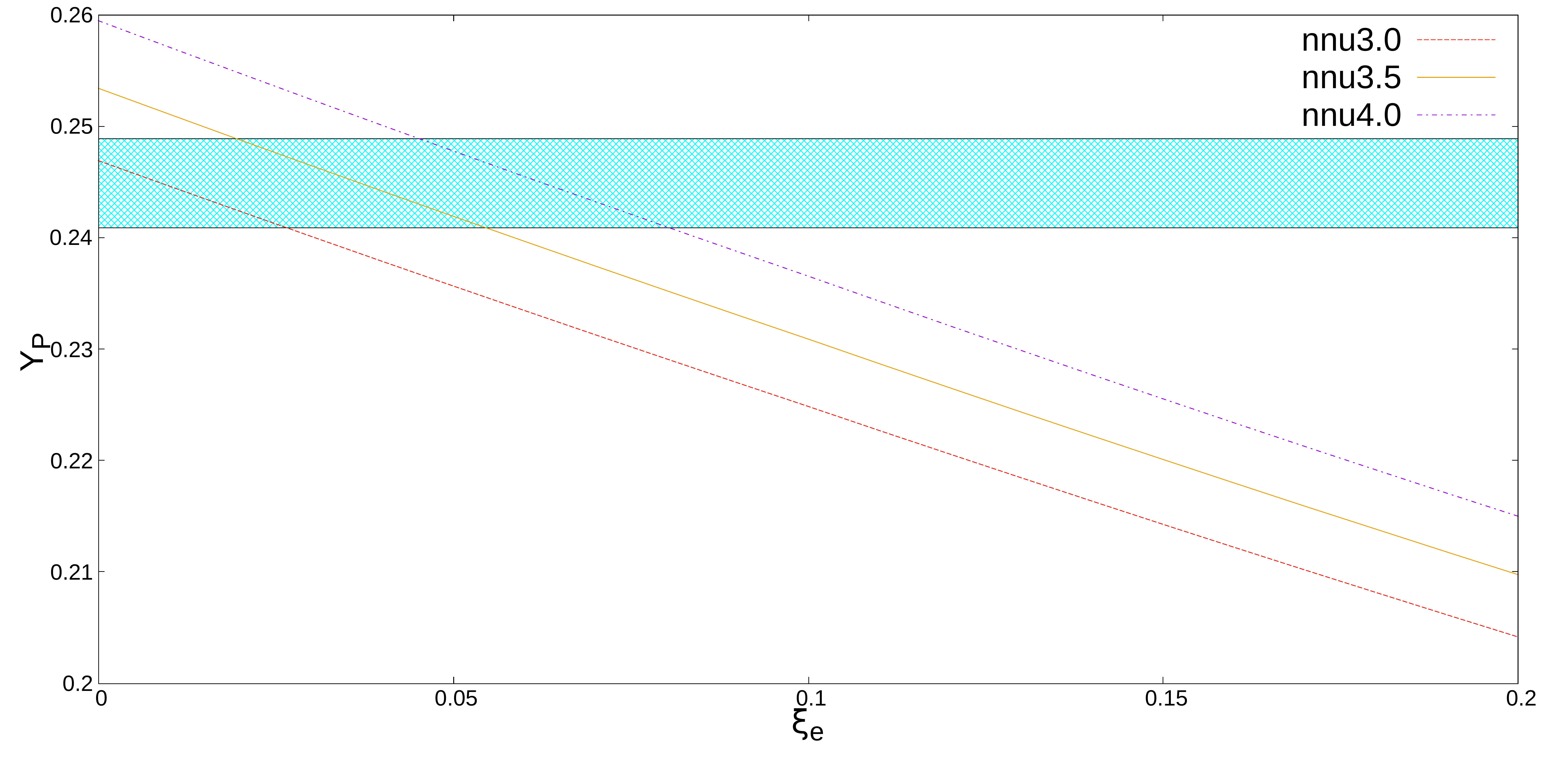}
\caption{
Effects of electron-type lepton asymmetry $\xi_{e}$ and the relativistic degrees
of freedom $N_{\mathrm{eff}}$ on the helium abundance $Y_{P}$. 
A larger $N_{\mathrm{eff}}$ leads to a larger $Y_{P}$, while a larger $\xi_{e}$
suppresses the resulting $Y_{P}$. The shading rectangle illustrates the value
of the $Y_{P}$ measurement: $Y_{P}=0.2449 \pm 0.0040$~\cite{Aver:2015iza}.
}
\label{Fig:PArthENoPE}
\end{figure}

We show, in Fig.~\ref{Fig:PArthENoPE}, a calculated $Y_P$ for various
$N_{\mathrm{eff}}$, $\xi_e$ and a fixed baryon asymmetry $\Omega_{b}h^{2}=0.223$. 
A larger $N_{\mathrm{eff}}$ leads to larger $Y_{P}$, while a larger
$\xi_{e}$ leads to smaller $Y_{P}$~\cite{Shiraishi:2009fu,Serpico:2005bc}.
The asymmetry $\xi_{e} \simeq 0.03$ s capable of canceling $\Delta N_{\mathrm{eff}} \simeq 0.5$ contribution to $Y_{P}$.
We find that $\xi_{e}\lesssim 0.03$ for $N_{\mathrm{eff}}=3.0$, 
$0.02\lesssim\xi_{e}\lesssim 0.06$ for $N_{\mathrm{eff}}=3.5$, 
and $0.05\lesssim\xi_{e}\lesssim 0.08$ for $N_{\mathrm{eff}}=4.0$ are
consistent with the $Y_{P}$ measurement.

Electron lepton asymmetry $\xi_{e}$ decreases slightly $D/H$.
In the fit to the CMB data, there is some correlation or parameter degeneracy
 between  $\Omega_{b}h^{2}$ and $\Delta N_{\mathrm{eff}}$.
The simultaneous increase of $\Delta N_{\mathrm{eff}}$ and $\Omega_{b}h^{2}$
 does not alter the resultant $D/H$ abundance~\cite{Cooke:2015yra,Consiglio:2017pot,Seto:2021xua}.

\begin{figure}[htbp]
\includegraphics[width=16cm]{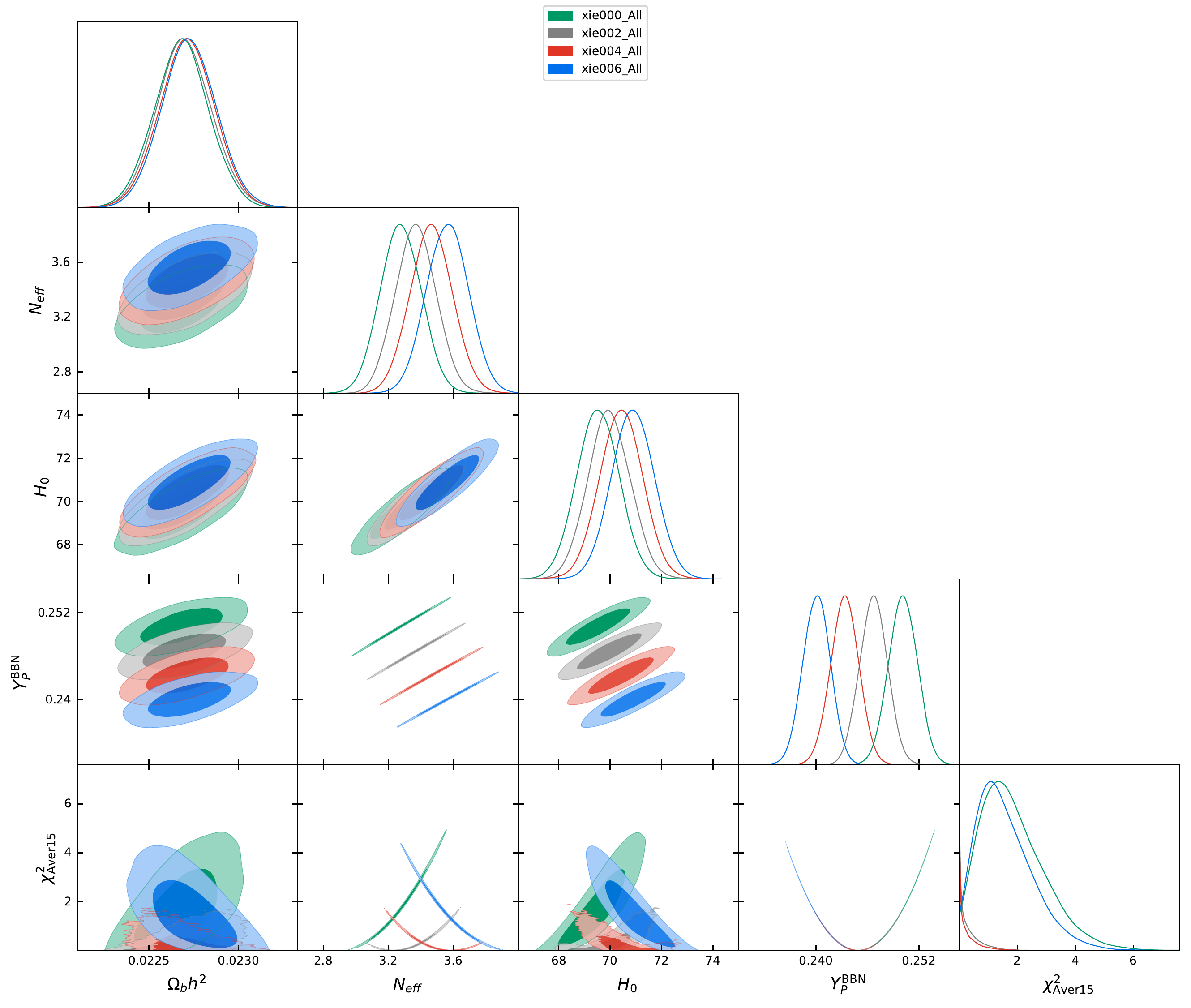}
\caption{Posterior distributions of $\Omega_{b}h^{2}$, $N_{\mathrm{eff}}$,
$H_{0}$, $Y_{P}$, and $\chi_{\mathrm{Aver}}^{2}$ on a $N_{\mathrm{eff}}$
model for several cases with differnet amount of lepton asymmetry.
This posterior have been derived for all datasets (CMB$+$BAO$+$JLA$+$BBN$+$R19). }
\label{Fig:MC1} 
\end{figure}

The Monte Carlo analysis has been carrid out using datasets (CMB $+$ BAO $+$ BBN $+$ JLA $+$ R19). 
The posteriors of parameters are summarized in Fig.~\ref{Fig:MC1}.
We show the results of four different neutrino degeneracy parameters $\xi_e = 0, 0.02, 0.04,$ and $0.06$.
As $\xi_e$ increases, a larger $N_\mathrm{eff}$ and, as the results, $H_0$ are allowed
 and the predicted $Y_P$ decreases.
With the change in $Y_P$, $\chi^2$ for the $Y_{P}$ measurement is minimized for $\xi_e \sim 0.04$.

The best-fit values of some quantities and the values of $\chi^2$ for the best-fit points
of $N_\mathrm{eff}$ models with several values of electron-type lepton degeneracy $\xi_e$
are summarized in Tab.~\ref{Table: chi^2}.
The total $\chi^{2}$ is also minimized at $\xi_{e} \simeq 0.04$ for two reasons. 
One is that, just mentioned above, light elements observations prefer $\xi_{e}\sim0.04$
($\chi_{{\rm Abund}}^{2}=\chi_{{\rm Cooke17}}^{2}+\chi_{{\rm Aver15}}^{2}$
at $\xi_{e}=0.04$ is about one-ninth of at $\xi_{e}=0.00$). 
The other reason is that the above datasets include R19, which indicates a larger $H_0$. 
When we introduce large asymmetry (e.g. $\xi_{e}=0.06$), a larger $N_\mathrm{eff}$ is required, 
then the Hubble tension could be relaxed greatly. 
However, such a too large $N_\mathrm{eff}$ is hardly compatible with both CMB and BBN. 
Therefore, we conclude that the most preferred electron
lepton asymmetry is $\xi_{e}=0.04$. This is consistent
with the previous works: $\xi=-0.002_{-0.111}^{+0.114}$ (95\%)
in Ref.~\cite{Oldengott:2017tzj}.
We obtain
\begin{align}
N_{\mathrm{eff}}=3.46 \pm 0.13,\quad  H_{0}=70.43 \pm 0.84 \,\mathrm{km/s/Mpc}, \quad \mathrm{for} \,\,\,\, \xi_{e}=0.04 \label{Eq:NeddH0xe004} \\
(68\%, \mathrm{Planck}+\mathrm{BAO}+\mathrm{Panthenon}+\mathrm{R19}+\mathrm{BBN}) . \nonumber 
\end{align}
%
%For comparison, we note 
%\begin{align}
%N_{\mathrm{eff}}=3.28 \pm 0.13,\quad H_{0}=69.53 \pm 0.82 \,\mathrm{km/s/Mpc}, \quad \mathrm{for} \,\,\,\, \xi_{e}=0.00  %\label{Eq:NeddH0xe000}\\
%(\mathrm{Planck}+\mathrm{BAO}+\mathrm{Panthenon}+\mathrm{R19}+{\mathrm{BBN}, 68\%)}. \nonumber
%\end{align}
%%
The values (\ref{Eq:NeddH0xe004}) show that the mean central value 
exceeds $H_0 = 70.0$ km/s/Mpc in $\xi_{e}=0.04$ cosmology.
The best-fit value is displayed in Tab.~\ref{Table: chi^2}.
By comparing the values (\ref{Eq:NeddH0xe004}) with the previous ones (\ref{Eq:DeltaN-w/o-xi}) for vanishing $\xi$ 
based on the same data set, we find favored values of $\Delta N_\mathrm{eff}$ is significantly shifted 
as $0.3 \lesssim \Delta N_\mathrm{eff} \lesssim 0.6 $ due to the lepton asymmetry.

\begin{table}[htbp]
\[
\begin{tabular}{lcccc}
 Parameter  &  \ensuremath{\xi_{e}}=0.00  &  \ensuremath{\xi_{e}}=0.02  &  \ensuremath{\xi_{e}}=0.04  &  \ensuremath{\xi_{e}}=0.06\\
\hline  \ensuremath{N_{\mathrm{eff}}}  &  3.243  &  3.313  &  3.455  &  3.634\\
 \ensuremath{H_{0}} [km/s/Mpc]  &  69.632  &  69.716  &  70.258  &  71.701\\
 \ensuremath{Y_{P}}  &  0.250  &  0.246  &  0.243  &  0.241 \\
\hline  
 \ensuremath{\chi_{{\rm Cooke17}}^{2}}  &  0.10  &  0.06  &  $1.5 \times 10^{-3}$  &  $6.6 \times 10^{-6}$ \\
 \ensuremath{\chi_{{\rm Aver15}}^{2}}  &  1.45  &  0.08  &  0.17  &  0.97 \\
 \ensuremath{\chi_{{\rm H074p03}}^{2}}  &  9.59  &  9.23  &  7.06  &  2.69 \\
 \ensuremath{\chi_{{\rm JLA}}^{2}}  &  1034.74  &  1034.74  &  1034.75  &  1034.81\\
 \ensuremath{\chi_{{\rm prior}}^{2}}  &  4.31  &  2.31  &  3.20  &  7.23 \\
 \ensuremath{\chi_{{\rm CMB}}^{2}}  &  2781.60  &  2783.90  &  2782.84  &  2783.71\\
 \ensuremath{\chi_{{\rm BAO}}^{2}}  &  5.80  &  5.41  &  5.38  &  6.57 \\
\hline  \ensuremath{\chi_{{\rm todal}}^{2}}  &  3837.61  &  3835.72  &  3833.39  &  3836.22 
\end{tabular}
\]
\caption{The best-fit $\chi^{2}$ for a $N_{\mathrm{eff}}$ model with and
without a lepton asymmetry.
Here, for reference, we note that the value of $\chi_{{\rm todal}}^2 = 3841.52$ for $\Lambda$CDM under 
the same data set has been derived~\cite{Seto:2021xua}. 
}
\label{Table: chi^2} 
\end{table}

\begin{figure}[htbp]
\includegraphics[width=16cm]{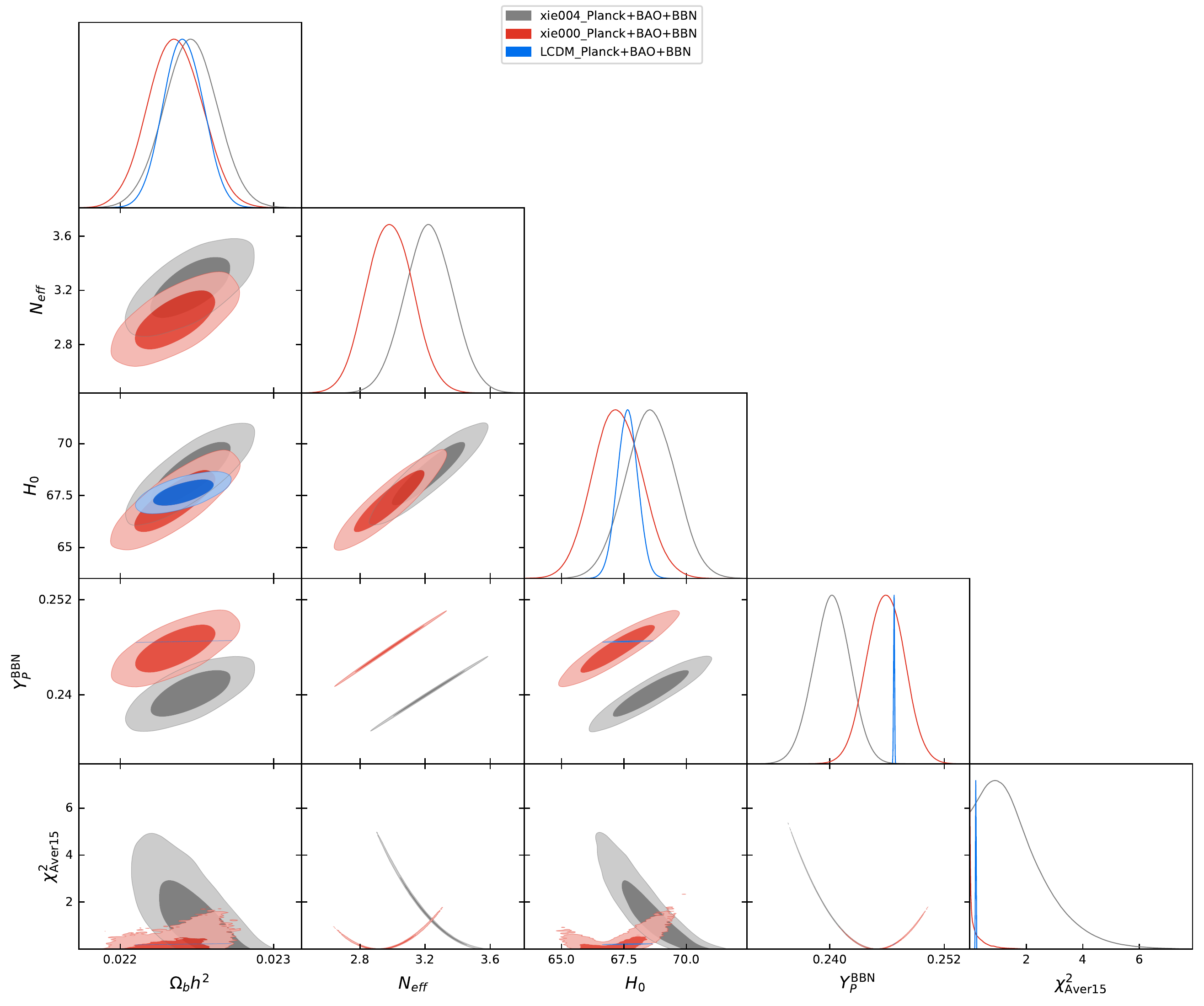}
\caption{Same as Fig.~\ref{Fig:MC1} for distant datasets (CMB$+$BAO$+$BBN) only.}
\label{Fig:MC2}
\end{figure}

We have also performed the analysis with the only distant data sets: Planck
$+$BAO$+$BBN at $\xi_{e}=0.04$, which is the best-fit point for the
combination of local and distant observations. 
The result is shown in Fig.~\ref{Fig:MC2} with $\Lambda$CDM and $N_\mathrm{eff}$ without any neutrino degeneracy
for comparison. We obtain
\begin{align}
N_{\mathrm{eff}}=3.22\pm 0.29,\quad  H_{0}=68.6\pm 2.0\,\mathrm{ km/s/Mpc}, \quad \mathrm{for} \,\,\,\, \xi_{e}=0.04 \\
(95\%, \mathrm{Planck}+\mathrm{BAO}+{\mathrm{BBN}}) . \nonumber
\end{align}
For comparison, we note 
\begin{align}
N_{\mathrm{eff}}=2.98_{-0.27}^{+0.28},\quad H_{0}=67.2_{-1.9}^{+2.0}\,\mathrm{ km/s/Mpc}, \quad \mathrm{for} \,\,\,\, \xi_{e}=0.00 \\
(95\%, \mathrm{Planck}+\mathrm{BAO}+{\mathrm{BBN}}), \nonumber
\end{align}
and quote $N_{\mathrm{eff}}=2.89\pm 0.29$ (95$\%$, Planck + BBN) from the Planck paper~\cite{Aghanim:2018eyx}. 
As shown in Fig.~\ref{Fig:MC2}, 
the value of $H_0 = 70.0$ km/s/Mpc lies in the uncertainty for the best-fit asymmetry $\xi_{e}=0.04$, 
even if we refer only distant observations.

\section{Summary}
\label{sec:summary}

We have studied cosmology with both the extra radiation parameterized by $N_\mathrm{eff}$
and a large lepton asymmetry, and examined the fit to various cosmological data.
In contrast to Ref.~\cite{Barenboim:2016lxv}, we consider the cases that an additional contribution, 
to $N_\mathrm{eff}$ beside that comes from the neutrino degeneracy $\xi$, 
such as hot axions and right-handed neutrinos although we do not need to specify species.

We have shown that a larger $N_\mathrm{eff}$ and $H_0$ are indicated for a larger neutrino degeneracy $\xi_e$.
At the best fit point with $\xi_{e}=0.04$ by taking all BBN, CMB, BAO and local measurements into account, 
the total $\chi^2$ of lepton asymmetric cosmology can be reduced by about $4$ from that of $N_\mathrm{eff}$ and 
by about $8$ from that of $\Lambda$CDM model.
We conclude that $\xi_{e}\simeq 0.04$ and $0.3 \lesssim \Delta N_\mathrm{eff} \lesssim 0.6 $
are favored to ameliorate the Hubble tension. 
The construction of such a model of particle physics able to reproduce these values would be worth examining.

Even if we analyze only the distant cosmological data of Planck, BAO and BBN, 
we can find a clear difference between the $N_\mathrm{eff}+\xi_e$ models and the $\Lambda$CDM.
For $\xi_{e}=0.04$, the data from Planck, BAO and BBN also infer significantly larger $H_0$ and $N_\mathrm{eff}$.
Then, $H_0 = 70.0$ km/s/Mpc is within the uncertainty, hence we might claim that, at least,
the tension between the Planck measurement and local $H_0$ measurements with TRGB calibration could be resolved 
by a large cosmological lepton asymmetry.

\section*{Acknowledgments}
This work is supported in part by the Japan Society for the Promotion
of Science (JSPS) KAKENHI Grants No.~19K03860, No.~19K03865 and No.~21H00060 (O.S.).

%%%%%%%%%%%%%%%%%%%%%%%%%%%%


\begin{thebibliography}{99}

\bibitem{Aghanim:2018eyx} 
N.~Aghanim \textit{et al.} [Planck],
%``Planck 2018 results. VI. Cosmological parameters,''
Astron. Astrophys. \textbf{641}, A6 (2020). %doi:10.1051/0004-6361/201833910
%[arXiv:1807.06209 [astro-ph.CO]].

\bibitem{Cyburt:2015mya} 
R.~H.~Cyburt, B.~D.~Fields, K.~A.~Olive and T.~H.~Yeh, 
%``Big Bang Nucleosynthesis: 2015,''
Rev. Mod. Phys. \textbf{88}, 015004 (2016).% doi:10.1103/RevModPhys.88.015004
%[arXiv:1505.01076 [astro-ph.CO]].

%%%%%%%%%%%%%%%%%%%%%%%%%%%%
%   Neff = 3.046
%%%%%%%%%%%%%%%%%%%%%%%%%%%%
%\cite{Mangano:2005cc}
\bibitem{Mangano:2005cc}
G.~Mangano, G.~Miele, S.~Pastor, T.~Pinto, O.~Pisanti and P.~D.~Serpico,
%``Relic neutrino decoupling including flavor oscillations,''
Nucl. Phys. B \textbf{729}, 221-234 (2005).
%doi:10.1016/j.nuclphysb.2005.09.041
%[arXiv:hep-ph/0506164 [hep-ph]].

%%%%%%%%%%%%%%%%%%%%%%%%%%%%
%   Neff calc
%%%%%%%%%%%%%%%%%%%%%%%%%%%%
%\cite{Escudero:2018mvt}
\bibitem{Escudero:2018mvt}
M.~Escudero,
%``Neutrino decoupling beyond the Standard Model: CMB constraints on the Dark Matter mass with a fast and precise $N_{\rm eff}$ evaluation,''
JCAP \textbf{02}, 007 (2019).
%doi:10.1088/1475-7516/2019/02/007
%[arXiv:1812.05605 [hep-ph]].
%\cite{Bennett:2019ewm}
\bibitem{Bennett:2019ewm}
J.~J.~Bennett, G.~Buldgen, M.~Drewes and Y.~Y.~Y.~Wong,
%``Towards a precision calculation of the effective number of neutrinos $N_{\rm eff}$ in the Standard Model I: the QED equation of state,''
JCAP \textbf{03}, 003 (2020).
%doi:10.1088/1475-7516/2020/03/003
%[arXiv:1911.04504 [hep-ph]].
%\cite{Escudero:2020dfa}
\bibitem{Escudero:2020dfa}
M.~Escudero Abenza,
%``Precision early universe thermodynamics made simple: $N_{\rm eff}$ and neutrino decoupling in the Standard Model and beyond,''
JCAP \textbf{05}, 048 (2020).
%doi:10.1088/1475-7516/2020/05/048
%[arXiv:2001.04466 [hep-ph]].
%\cite{Akita:2020szl}
\bibitem{Akita:2020szl}
K.~Akita and M.~Yamaguchi,
%``A precision calculation of relic neutrino decoupling,''
JCAP \textbf{08}, 012 (2020).
%doi:10.1088/1475-7516/2020/08/012
%[arXiv:2005.07047 [hep-ph]].
%\cite{Bennett:2020zkv}
\bibitem{Bennett:2020zkv}
J.~J.~Bennett, G.~Buldgen, P.~F.~De Salas, M.~Drewes, S.~Gariazzo, S.~Pastor and Y.~Y.~Y.~Wong,
%``Towards a precision calculation of $N_{\rm eff}$ in the Standard Model II: Neutrino decoupling in the presence of flavour oscillations and finite-temperature QED,''
[arXiv:2012.02726 [hep-ph]].


\bibitem{Aiola:2020azj} 
S.~Aiola \textit{et al.} [ACT], %``The Atacama Cosmology Telescope: DR4 Maps and Cosmological Parameters,''
JCAP \textbf{12}, 047 (2020).
% doi:10.1088/1475-7516/2020/12/047 [arXiv:2007.07288
%[astro-ph.CO]].

\bibitem{Addison:2017fdm} 
G.~E.~Addison, D.~J.~Watts, C.~L.~Bennett, M.~Halpern, G.~Hinshaw and J.~L.~Weiland, 
%``Elucidating $\Lambda$CDM: Impact of Baryon Acoustic Oscillation Measurements on the Hubble Constant Discrepancy,''
Astrophys. J. \textbf{853}, no.2, 119 (2018).% doi:10.3847/1538-4357/aaa1ed
%[arXiv:1707.06547 [astro-ph.CO]].

%\cite{Schoneberg:2019wmt}
\bibitem{Schoneberg:2019wmt}
N.~Sch\"oneberg, J.~Lesgourgues and D.~C.~Hooper,
%``The BAO+BBN take on the Hubble tension,''
JCAP \textbf{10}, 029 (2019).
%doi:10.1088/1475-7516/2019/10/029
%[arXiv:1907.11594 [astro-ph.CO]].

%\cite{Riess:2018uxu}
\bibitem{Riess:2018uxu}
A.~G.~Riess \textit{et al.},
% S.~Casertano, W.~Yuan, L.~Macri, J.~Anderson, J.~W.~MacKenty, J.~Bradley Bowers, K.~I.~Clubb, A.~V.~Filippenko and D.~O.~Jones, \textit{et al.}
%``New Parallaxes of Galactic Cepheids from Spatially Scanning the Hubble Space Telescope: Implications for the Hubble Constant,''
Astrophys. J. \textbf{855}, no.2, 136 (2018).
%doi:10.3847/1538-4357/aaadb7
%[arXiv:1801.01120 [astro-ph.SR]].
\bibitem{Riess:2019cxk} 
A.~G.~Riess, S.~Casertano, W.~Yuan, L.~M.~Macri and D.~Scolnic, %``Large Magellanic Cloud Cepheid Standards Provide a 1% Foundation for the Determination of the Hubble Constant and Stronger Evidence for Physics beyond $\Lambda$CDM,''
Astrophys. J. \textbf{876}, no.1, 85 (2019). % doi:10.3847/1538-4357/ab1422 
%[arXiv:1903.07603 [astro-ph.CO]].

\bibitem{Wong:2019kwg} 
K.~C.~Wong, S.~H.~Suyu, G.~C.~F.~Chen, C.~E.~Rusu, M.~Millon, D.~Sluse, V.~Bonvin, C.~D.~Fassnacht,
S.~Taubenberger and M.~W.~Auger, \textit{et al.} %``H0LiCOW \textendash{} XIII. A 2.4 per cent measurement of H0 from lensed quasars: 5.3\ensuremath{\sigma} tension between early- and late-Universe probes,''
Mon. Not. Roy. Astron. Soc. \textbf{498}, no.1, 1420-1439 (2020).
% doi:10.1093/mnras/stz3094
%[arXiv:1907.04869 [astro-ph.CO]].

%\cite{Freedman:2019jwv}
\bibitem{Freedman:2019jwv}
W.~L.~Freedman, B.~F.~Madore, D.~Hatt, T.~J.~Hoyt, I.~S.~Jang, R.~L.~Beaton, C.~R.~Burns, M.~G.~Lee, A.~J.~Monson and J.~R.~Neeley, \textit{et al.}
%``The Carnegie-Chicago Hubble Program. VIII. An Independent Determination of the Hubble Constant Based on the Tip of the Red Giant Branch,''
Astrophys. J. \textbf{882}, 34 (2019).
%doi:10.3847/1538-4357/ab2f73
%[arXiv:1907.05922 [astro-ph.CO]].

%\cite{DiValentino:2021izs}
\bibitem{DiValentino:2021izs}
E.~Di Valentino, O.~Mena, S.~Pan, L.~Visinelli, W.~Yang, A.~Melchiorri, D.~F.~Mota, A.~G.~Riess and J.~Silk,
%``In the Realm of the Hubble tension $-$ a Review of Solutions,''
[arXiv:2103.01183 [astro-ph.CO]].

\bibitem{Scolnic:2017caz} 
D.~M.~Scolnic, D.~O.~Jones, A.~Rest, Y.~C.~Pan, R.~Chornock, R.~J.~Foley, M.~E.~Huber, R.~Kessler,
G.~Narayan and A.~G.~Riess, \textit{et al.} 
%``The Complete Light-curve Sample of Spectroscopically Confirmed SNe Ia from Pan-STARRS1 and Cosmological Constraints from the Combined Pantheon Sample,''
Astrophys. J. \textbf{859}, no.2, 101 (2018). 
%doi:10.3847/1538-4357/aab9bb
%[arXiv:1710.00845 [astro-ph.CO]].

\bibitem{Seto:2021xua} 
O.~Seto and Y.~Toda, 
%``Comparing early dark energy and extra radiation solutions to the Hubble tension with BBN,''
[arXiv:2101.03740 [astro-ph.CO]].

\bibitem{DEramo:2018vss} 
F.~D'Eramo, R.~Z.~Ferreira, A.~Notari and J.~L.~Bernal, 
%``Hot Axions and the $H_0$ tension,''
JCAP \textbf{11}, 014 (2018). 
% doi:10.1088/1475-7516/2018/11/014 [arXiv:1808.07430[hep-ph]].

\bibitem{Escudero:2019gzq} 
M.~Escudero, D.~Hooper, G.~Krnjaic and M.~Pierre, 
%``Cosmology with A Very Light L$_{\mu}$ â L$_{\tau}$ Gauge Boson,''
JHEP \textbf{03}, 071 (2019). 
% doi:10.1007/JHEP03(2019)071 [arXiv:1901.02010 [hep-ph]].

%%%%%%%%%%%%%%%%%%%%%%%%%%%%
%   lepton asymm on BBN
%%%%%%%%%%%%%%%%%%%%%%%%%%%%
%
\bibitem{BeaudetGoret}
G.~Beaudet and P.~Goret,
%Leptonic numbers and the neutron to proton ratio in the hot Big Bang model. 
Astronomy and Astrophysics, \textbf{49}, no. 3, 415-419 (1976).
%%%%%%%%%%%
\bibitem{Lesgourgues:1999wu} 
J.~Lesgourgues and S.~Pastor, 
%``Cosmological implications of a relic neutrino asymmetry,''
Phys. Rev. D \textbf{60}, 103521 (1999).
% doi:10.1103/PhysRevD.60.103521
%[arXiv:hep-ph/9904411 [hep-ph]].
%%%%%%%%%%%
%\cite{Serpico:2005bc}
\bibitem{Serpico:2005bc} 
P.~D.~Serpico and G.~G.~Raffelt, 
%``Lepton asymmetry and primordial nucleosynthesis in the era of precision cosmology,''
Phys. Rev. D \textbf{71}, 127301 (2005).
% doi:10.1103/PhysRevD.71.127301
%[arXiv:astro-ph/0506162 [astro-ph]]. 
%%%%%%%%%%%
\bibitem{Shiraishi:2009fu} 
M.~Shiraishi, K.~Ichikawa, K.~Ichiki, N.~Sugiyama and M.~Yamaguchi, 
%``Constraints on neutrino masses from WMAP5 and BBN in the lepton asymmetric universe,''
JCAP \textbf{07}, 005 (2009).% doi:10.1088/1475-7516/2009/07/005 [arXiv:0904.4396
%[astro-ph.CO]].
%%%%%%%%%%%
%\cite{Kirilova:2013aja}
\bibitem{Kirilova:2013aja} 
D.~Kirilova, 
%``Lepton Asymmetry and Neutrino Oscillations Interplay,''
Hyperfine Interact. \textbf{215}, no.1-3, 111-118 (2013).
% doi:10.1007/s10751-013-0790-0
%[arXiv:1302.2923 [astro-ph.CO]].
%%%%%%%%%%%
%\cite{Caramete:2013bua}
\bibitem{Caramete:2013bua} 
A.~Caramete and L.~A.~Popa, 
%``Cosmological evidence for leptonic asymmetry after Planck,''
JCAP \textbf{02}, 012 (2014).
% doi:10.1088/1475-7516/2014/02/012 [arXiv:1311.3856[astro-ph.CO]].
%%%%%%%%%%%
\bibitem{Yang:2018oqg} 
C.~T.~Yang, J.~Birrell and J.~Rafelski,
%``Lepton Number and Expansion of the Universe,''
[arXiv:1812.05157 [hep-ph]].
%%%%%%%%%%%
%\cite{Barenboim:2016lxv}
\bibitem{Barenboim:2016lxv}
G.~Barenboim, W.~H.~Kinney and W.~I.~Park,
%``Flavor versus mass eigenstates in neutrino asymmetries: implications for cosmology,''
Eur. Phys. J. C \textbf{77}, no.9, 590 (2017).
%doi:10.1140/epjc/s10052-017-5147-4
%[arXiv:1609.03200 [astro-ph.CO]].
%%%%%%%%%%%
%\cite{Gelmini:2020ekg}
\bibitem{Gelmini:2020ekg}
G.~B.~Gelmini, M.~Kawasaki, A.~Kusenko, K.~Murai and V.~Takhistov,
%``Big Bang Nucleosynthesis constraints on sterile neutrino and lepton asymmetry of the Universe,''
JCAP \textbf{09}, 051 (2020).
%doi:10.1088/1475-7516/2020/09/051
%[arXiv:2005.06721 [hep-ph]].

%%%%%%%%%%%%%%%%%%%%%%%%%%%%
%   lepton asymm. mechanism
%%%%%%%%%%%%%%%%%%%%%%%%%%%%
%\cite{Foot:1995qk}
\bibitem{Foot:1995qk}
R.~Foot, M.~J.~Thomson and R.~R.~Volkas,
%``Large neutrino asymmetries from neutrino oscillations,''
Phys. Rev. D \textbf{53}, R5349-R5353 (1996).
%doi:10.1103/PhysRevD.53.R5349
%[arXiv:hep-ph/9509327 [hep-ph]].
%\cite{Shi:1996ic}
\bibitem{Shi:1996ic}
X.~D.~Shi,
%``Chaotic amplification of neutrino chemical potentials by neutrino oscillations in big bang nucleosynthesis,''
Phys. Rev. D \textbf{54}, 2753-2760 (1996).
%doi:10.1103/PhysRevD.54.2753
%[arXiv:astro-ph/9602135 [astro-ph]].
%\cite{Casas:1997gx}
\bibitem{Casas:1997gx}
A.~Casas, W.~Y.~Cheng and G.~Gelmini,
%``Generation of large lepton asymmetries,''
Nucl. Phys. B \textbf{538}, 297-308 (1999).
%doi:10.1016/S0550-3213(98)00606-3
%[arXiv:hep-ph/9709289 [hep-ph]].
%\cite{McDonald:1999in}
\bibitem{McDonald:1999in}
J.~McDonald,
%``Naturally large cosmological neutrino asymmetries in the MSSM,''
Phys. Rev. Lett. \textbf{84}, 4798-4801 (2000).
%doi:10.1103/PhysRevLett.84.4798
%[arXiv:hep-ph/9908300 [hep-ph]].
%\cite{Kawasaki:2002hq}
\bibitem{Kawasaki:2002hq}
M.~Kawasaki, F.~Takahashi and M.~Yamaguchi,
%``Large lepton asymmetry from Q balls,''
Phys. Rev. D \textbf{66}, 043516 (2002).
%doi:10.1103/PhysRevD.66.043516
%[arXiv:hep-ph/0205101 [hep-ph]].
%\cite{Yamaguchi:2002vw}
\bibitem{Yamaguchi:2002vw}
M.~Yamaguchi,
%``Generation of cosmological large lepton asymmetry from a rolling scalar field,''
Phys. Rev. D \textbf{68}, 063507 (2003).
%doi:10.1103/PhysRevD.68.063507
%[arXiv:hep-ph/0211163 [hep-ph]].
%\cite{Takahashi:2003db}
\bibitem{Takahashi:2003db}
F.~Takahashi and M.~Yamaguchi,
%``Spontaneous baryogenesis in flat directions,''
Phys. Rev. D \textbf{69}, 083506 (2004).
%doi:10.1103/PhysRevD.69.083506
%[arXiv:hep-ph/0308173 [hep-ph]].
%\cite{Shaposhnikov:2008pf}
\bibitem{Shaposhnikov:2008pf}
M.~Shaposhnikov,
%``The nuMSM, leptonic asymmetries, and properties of singlet fermions,''
JHEP \textbf{08}, 008 (2008).
%doi:10.1088/1126-6708/2008/08/008
%[arXiv:0804.4542 [hep-ph]].


\bibitem{Aver:2015iza} 
E.~Aver, K.~A.~Olive and E.~D.~Skillman,
%``The effects of He I \ensuremath{\lambda}10830 on helium abundance determinations,''
JCAP \textbf{07}, 011 (2015). %doi:10.1088/1475-7516/2015/07/011
%[arXiv:1503.08146 [astro-ph.CO]].


%%%%%%%%%%%%%%%%%%%%%%%%%%%%
%   lepton asymm with osc.
%%%%%%%%%%%%%%%%%%%%%%%%%%%%
%
%\cite{Dolgov:2002ab}
\bibitem{Dolgov:2002ab}
A.~D.~Dolgov, S.~H.~Hansen, S.~Pastor, S.~T.~Petcov, G.~G.~Raffelt and D.~V.~Semikoz,
%``Cosmological bounds on neutrino degeneracy improved by flavor oscillations,''
Nucl. Phys. B \textbf{632}, 363-382 (2002).
%doi:10.1016/S0550-3213(02)00274-2
%[arXiv:hep-ph/0201287 [hep-ph]].
%\cite{Wong:2002fa}
\bibitem{Wong:2002fa}
Y.~Y.~Y.~Wong,
%``Analytical treatment of neutrino asymmetry equilibration from flavor oscillations in the early universe,''
Phys. Rev. D \textbf{66}, 025015 (2002).
%doi:10.1103/PhysRevD.66.025015
%[arXiv:hep-ph/0203180 [hep-ph]].
%\cite{Abazajian:2002qx}
\bibitem{Abazajian:2002qx}
K.~N.~Abazajian, J.~F.~Beacom and N.~F.~Bell,
%``Stringent Constraints on Cosmological Neutrino Antineutrino Asymmetries from Synchronized Flavor Transformation,''
Phys. Rev. D \textbf{66}, 013008 (2002).
%doi:10.1103/PhysRevD.66.013008
%[arXiv:astro-ph/0203442 [astro-ph]].
%\cite{Pastor:2008ti}
\bibitem{Pastor:2008ti}
S.~Pastor, T.~Pinto and G.~G.~Raffelt,
%``Relic density of neutrinos with primordial asymmetries,''
Phys. Rev. Lett. \textbf{102}, 241302 (2009).
%doi:10.1103/PhysRevLett.102.241302
%[arXiv:0808.3137 [astro-ph]].
%\cite{Barenboim:2016shh}
\bibitem{Barenboim:2016shh}
G.~Barenboim, W.~H.~Kinney and W.~I.~Park,
%``Resurrection of large lepton number asymmetries from neutrino flavor oscillations,''
Phys. Rev. D \textbf{95}, no.4, 043506 (2017).
%doi:10.1103/PhysRevD.95.043506
%[arXiv:1609.01584 [hep-ph]].


\bibitem{Ma:1995ey} 
C.~P.~Ma and E.~Bertschinger, 
%``Cosmological perturbation theory in the synchronous and conformal Newtonian gauges,''
Astrophys. J. \textbf{455}, 7-25 (1995).% doi:10.1086/176550 [arXiv:astro-ph/9506072[astro-ph]].

\bibitem{Lewis:2002ah} 
A.~Lewis and S.~Bridle, 
%``Cosmological parameters from CMB and other data: A Monte Carlo approach,''
Phys. Rev. D \textbf{66}, 103511 (2002). %doi:10.1103/PhysRevD.66.103511
%[arXiv:astro-ph/0205436 [astro-ph]].

\bibitem{Consiglio:2017pot} 
R.~Consiglio, P.~F.~de Salas, G.~Mangano, G.~Miele, S.~Pastor and O.~Pisanti, 
%``PArthENoPE reloaded,''
Comput. Phys. Commun. \textbf{233}, 237-242 (2018).
% doi:10.1016/j.cpc.2018.06.022 
%[arXiv:1712.04378 [astro-ph.CO]].

\bibitem{Aghanim:2018oex} 
N.~Aghanim \textit{et al.} [Planck],
%``Planck 2018 results. VIII. Gravitational lensing,''
Astron. Astrophys. \textbf{641}, A8 (2020). 
%doi:10.1051/0004-6361/201833886
%[arXiv:1807.06210 [astro-ph.CO]].

\bibitem{Beutler:2011hx} 
F.~Beutler, C.~Blake, M.~Colless, D.~H.~Jones, L.~Staveley-Smith, L.~Campbell, Q.~Parker, W.~Saunders and F.~Watson,
%``The 6dF Galaxy Survey: Baryon Acoustic Oscillations and the Local Hubble Constant,''
Mon. Not. Roy. Astron. Soc. \textbf{416}, 3017-3032 (2011).

\bibitem{Ross:2014qpa} 
A.~J.~Ross, L.~Samushia, C.~Howlett, W.~J.~Percival, A.~Burden and M.~Manera, 
%``The clustering of the SDSS DR7 main Galaxy sample \textendash{} I. A 4 per cent distance measure at $z = 0.15$,''
Mon. Not. Roy. Astron. Soc. \textbf{449}, no.1, 835-847 (2015). 
%doi:10.1093/mnras/stv154
%[arXiv:1409.3242 [astro-ph.CO]].

\bibitem{Alam:2016hwk} 
S.~Alam \textit{et al.} [BOSS], 
%``The clustering of galaxies in the completed SDSS-III Baryon Oscillation Spectroscopic Survey: cosmological analysis of the DR12 galaxy sample,''
Mon. Not. Roy. Astron. Soc. \textbf{470}, no.3, 2617-2652 (2017).
%doi:10.1093/mnras/stx721
%[arXiv:1607.03155 [astro-ph.CO]].

\bibitem{Cooke:2017cwo} 
R.~J.~Cooke, M.~Pettini and C.~C.~Steidel,
%``One Percent Determination of the Primordial Deuterium Abundance,''
Astrophys. J. \textbf{855}, no.2, 102 (2018). 
%doi:10.3847/1538-4357/aaab53
%[arXiv:1710.11129 [astro-ph.CO]].

%\cite{Cooke:2015yra}
\bibitem{Cooke:2015yra}
R.~J.~Cooke,
%``Big Bang Nucleosynthesis and the Helium Isotope Ratio,''
Astrophys. J. Lett. \textbf{812}, no.1, L12 (2015).
%doi:10.1088/2041-8205/812/1/L12
%[arXiv:1510.02801 [astro-ph.CO]].

\bibitem{Oldengott:2017tzj} 
I.~M.~Oldengott and D.~J.~Schwarz,
%``Improved constraints on lepton asymmetry from the cosmic microwave background,''
EPL \textbf{119}, no.2, 29001 (2017).
% doi:10.1209/0295-5075/119/29001
%[arXiv:1706.01705 [astro-ph.CO]].

%\cite{Bonilla:2018nau}



\end{thebibliography}
\end{document}